\documentclass[prl,twocolumn,twoside,superscriptaddress,showpacs,floatfix]{revtex4}
\usepackage{dcolumn}
\usepackage{amsmath}
\usepackage{graphicx}

\begin{document}
\bibliographystyle{apsrev}

\preprint{Santa Fe Institute Working Paper 02-04-017}

\title{Coupled Replicator Equations for the\\
Dynamics of Learning in Multiagent Systems}

\author{Yuzuru Sato}
\email{ysato@bdc.brain.riken.go.jp}
\affiliation{Brain Science Institute, Institute of Physical and
Chemical Research (RIKEN), 2-1 Hirosawa, 
Saitama 351-0198, Japan}
\affiliation{Santa Fe Institute, 1399 Hyde Park Road, Santa Fe, NM 87501}

\author{James P. Crutchfield}
\email{chaos@santafe.edu}
\affiliation{Santa Fe Institute, 1399 Hyde Park Road, Santa Fe, NM 87501}

\date{\today}

\begin{abstract}
Starting with a group of reinforcement-learning agents we derive coupled
replicator equations that describe the dynamics of collective learning in
multiagent systems. We show that, although agents model their environment
in a self-interested way without sharing knowledge, a game dynamics emerges
naturally through environment-mediated interactions. An application to
rock-scissors-paper game interactions shows that the collective learning
dynamics exhibits a diversity of competitive and cooperative behaviors.
These include quasiperiodicity, stable limit cycles, intermittency, and
deterministic chaos---behaviors that should be expected in heterogeneous
multiagent systems described by the general replicator equations we derive. 
\end{abstract}
\pacs{05.45.-a, 02.50.Le, 87.23.-n
~~~~~~~~~~~~~~~~~~~~Santa Fe Institute Working Paper 02-04-017}
\maketitle

Adaptive behavior in multiagent systems is an important interdisciplinary
topic that appears in various guises in many fields, including biology
\cite{Cama01a}, computer science \cite{Simo96}, economics \cite{Youn98a}, 
and cognitive science \cite{Hutc96a}. One of the key common questions is
how and whether a group of intelligent agents truly engages in collective
behaviors that are more functional than individuals acting alone. 

Suppose that many agents interact with an environment and each independently
builds a model from its sensory stimuli. In this simple type of coupled
multiagent system, collective learning (if it occurs) is a dynamical behavior
driven by agents' environment-mediated interaction \cite{Ros87,Tai99}.
Here we show that the collective dynamics in multiagent systems, in which
agents use reinforcement learning \cite{Sutt98a}, can be modeled using a
generalized form of coupled replicator equations.

While replicator dynamics were introduced originally for evolutionary game
theory \cite{Tay78}, the relationship between reinforcement learning and
replicator equations has been developed only recently \cite{Bor97}. Here, we
extend these considerations to multiagent systems, introducing the theory
behind a previously reported game-theoretic model \cite{Sat02}. We show that
replicator dynamics emerges as a special case of the continuous-time limit
for multiagent reinforcement learning systems. The overall approach, though,
establishes a general framework for dynamical-systems analyses of adaptive
behavior in collectives.

Notably, in learning with perfect memory, our model reduces to the form
of a multipopulation replicator equation introduced in Ref. \cite{Tay79}.
For two agents with perfect memory interacting via a zero-sum
rock-scissors-paper game the dynamics exhibits Hamiltonian chaos \cite{Sat02}.
In contrast, as we show here, with memory decay multiagent systems generally
become dissipative and display the full range of nonlinear dynamical behaviors,
including limit cycles, intermittency, and deterministic chaos.

Our multiagent model begins with simple reinforcement learning
agents. To simplify the development, we assume that there are
two such agents $X$ and $Y$ that at each time step take one of
$N$ actions: $i = 1, \ldots, N$. Let the probability for $X$ to chose action
$i$ be $x_i(n)$ and $y_i(n)$ for $Y$, where $n$ is the number of the learning
iterations from the initial state at $n = 0$. The agents' choice
distributions at time $n$ are ${\bf x}(n)=(x_1(n), \ldots,x_N(n))$ and  
${\bf y}(n) = (y_1(n), \ldots, y_N(n))$, with
$\Sigma_i x_i(n) = \Sigma_i y_i(n) = 1$.

Let $R_{ij}^X$ and $R_{ij}^Y$ denote the reward for $X$ taking action $i$
and $Y$ action $j$ at step $n$, respectively. Given these actions, $X$'s and
$Y$'s memories, $Q_i^X (n)$ and $Q_i^Y (n)$, of the past benefits
from their actions are governed by
\begin{eqnarray}
Q_i^X (n+1)-Q_i^X(n)& = & R_{ij}^X - \alpha_X Q_i^X (n) ~{\rm and}\\ \nonumber
Q_i^Y (n+1)-Q_i^Y(n)& = & R_{ij}^Y - \alpha_Y Q_i^Y (n) ~,
\label{MemoryUpdate}
\end{eqnarray}
where $\alpha_X, \alpha_Y \in [0,1)$ control each agent's memory decay rate
and $Q_i^X (0) = Q_i^Y(0) = 0$. The agents chose their next actions according
to the $Q$'s, updating their choice distributions as follows: 
\begin{equation}
x_i (n) = \frac{e^{\beta_X Q_i^X (n)}} {\Sigma_j e^{\beta_X Q_j^X (n)}}
~{\rm and}~
y_i (n) = \frac{e^{\beta_Y Q_i^Y (n)}} {\Sigma_j e^{\beta_Y Q_j^Y (n)}}, 
\label{Vector}
\end{equation}
where $\beta_X, \beta_Y\in [0,\infty]$ control the learning sensitivity:
how much the current choice distributions are affected by past rewards. 
Using Eq. (\ref{Vector}), the dynamic governing the change in agent state 
is given by:  
\begin{equation}
x_i (n+1) = \frac{x_i (n) e^{\beta_X (Q_i^X(n+1)-Q_i^X(n))}}
  {\Sigma_k x_j (n) e^{\beta_X (Q_k^X(n+1)-Q_k^X (n))}} ~,
\label{VectorUpdate} 
\end{equation}
and similarly for $y_i (n+1)$.

Consider the continuous-time limit corresponding to agents performing a large
number of actions (iterates of Eqs. (1)) for each choice-distribution
update (iterates of Eq. (\ref{VectorUpdate})). In this case, we have two 
different time scales---that for agent-agent interactions and for learning.
We assume that the learning dynamics is very slow compared to interactions
and so $\bf x$ and $\bf y$ are essentially constant during the latter.
Then, based on Eq. (\ref{VectorUpdate}), continuous-time learning for agent
$X$ is governed by
\begin{equation} 
\dot{x}_i = \beta_X x_i (\dot{Q}_i^X - \Sigma_j \dot{Q}^X_j x_j) ~,
\label{eq:rep}
\end{equation}
and for the dynamic governing memory updates we have 
\begin{equation}
\dot{Q}_i^X = R_i^X - \alpha_X Q_i^X ~, 
\label{eq:q}
\end{equation}
where $R_i^X$ is the reward for $X$ choosing action $i$, averaged over $Y$'s
actions during the time interval between learning updates. Putting together
Eqs. (\ref{Vector}), (\ref{eq:rep}), and (\ref{eq:q}) one finds
\begin{equation} 
\frac{\dot{x_i}}{x_i} 
  = \beta_X[R_i^X-\Sigma_j x_{ij}R_j^X] +\alpha_X I^X_i~, 
\label{eq:crex1}
\end{equation}
where $I^X_i \equiv \Sigma_j x_j \log ({x_j}/{x_i})$ represents the effect
of memory with decay parameter $\alpha_X$. (The continuous-time dynamic of $Y$
follows in a similar manner.) Eq. (\ref{eq:crex1}), extended to account for
any number of agents and actions, constitutes our general model for
reinforcement-learning multiagent systems.

Simplifying again, assume a fixed relationship between pairs $(i,j)$ of $X$'s
and $Y$'s actions and between rewards for both agents: $R_{ij}^X = a_{ij}$ and
$R_{ij}^Y = b_{ij}$. Assume further that $\bf x$ and $\bf y$ are independently
distributed, then the time-average rewards for $X$ and $Y$ become
\begin{equation}
  R_i^X = \Sigma_j a_{ij} y_j ~{\rm and}~ R_i^Y =\Sigma_j b_{ij} x_j~,
\label{eq:r}
\end{equation}
In this restricted case, the continuous-time dynamic is:
\begin{eqnarray} 
\frac{\dot{x}_i}{x_i} & = & \beta_X [ (A{\bf y})_i-{\bf x}\cdot A{\bf y}]
  + \alpha_X I^X_i ~, \nonumber \\
\frac{\dot{y}_i}{y_i} & = & \beta_Y [ (B{\bf x})_i-{\bf y}\cdot B{\bf x}]
  + \alpha_Y I^Y_i ~,
\label{eq:crex}
\end{eqnarray}
where $(A)_{ij} = a_{ij}$ and $(B)_{ij} = b_{ij}$, $(A{\bf x})_i$ is the
$i$th element of the vector $A{\bf x}$, and $\beta_X$ and $\beta_Y$ control
the time-scale of each agent's learning.  

We can regard $A$ and $B$ as $X$'s and $Y$'s game-theoretic payoff matrices
for action $i$ against opponent's action $j$ \footnote{Eqs. (\ref{eq:r})
specify the von Neumann-Morgenstern utility (J. von Neumann and O. Morgenstern,
{\it Theory of Games and Economic Behavior}, (Princeton University Press,
1944)).}. In contrast with game theory, which assumes agents have exact
knowledge of the game structure and of other agent's strategies,
reinforcement-learning agents have no knowledge of a ``game'' in which
they are playing, only a myopic model of the
environment---other agent(s)---given implicitly via the rewards they receive.
Nonetheless, a game dynamics emerges---via $R^X$ and $R^Y$
in Eq. (\ref{eq:crex1})---as a description of the collective's {\em global
behavior}.

Given the basic equations of motion for the reinforcement-learning multiagent
system (Eq. (\ref{eq:crex})), one becomes interested in, on the one hand, the
time evolution of each agent's state vector in the simplices
${\bf x}\in \Delta_X$ and ${\bf y} \in \Delta_Y$ and, on the other, the
dynamics in the higher-dimensional {\em collective} simplex
$({\bf x},{\bf y}) \in \Delta_X \times \Delta_Y$.
Following Ref. \cite{Hof96}, we transform from
$({\bf x}, {\bf y}) \in \Delta_X \times \Delta_Y$ to
${\bf U}=({\bf u}, {\bf v})\in{\bf R}^{2(N-1)}$ with 
${\bf u}=(u_1, \ldots, u_{N-1})$ and 
${\bf v}=(v_1, \ldots, v_{N-1})$, where
$u_i = \log (x_{i+1}/x_1) ~{\rm and}~ v_i=\log (y_{i+1}/y_1),
 ~(i=1, \ldots, N-1)$.
The result is a new version of our simplified model (Eqs. (\ref{eq:crex})),
useful both for numerical stability during simulation and also
for analysis in certain limits:
\begin{eqnarray}
\dot{u}_i & = & \beta_X
  \frac{\Sigma_j \tilde{a}_{ij} e^{v_j} + \tilde{a}_{i1}}
  {1 + \Sigma_j e^{v_j}} - \alpha_X u_i ~{\rm and} \nonumber \\
\dot{v}_i & = & \beta_Y
  \frac{\Sigma_j \tilde{b}_{ij} e^{u_j} + \tilde{b}_{i1}}
  {1 + \Sigma_j e^{u_j}} - \alpha_Y v_i ~,
\label{TransformedEoM}
\end{eqnarray}
where $\tilde{a}_{ij} = a_{i+1,j}-a_{1,j}$ and
$\tilde{b}_{ij} = b_{i+1,j}-b_{1,j}$.
Since the dissipation rate $\gamma$ in $\bf U$ is
\begin{equation}
\gamma = \Sigma_i\frac{\partial \dot{u}_i}{\partial u_i}
  + \Sigma_j\frac{\partial \dot{v}_j}{\partial v_j}
  = -(N-1) (\alpha_X + \alpha_Y), 
\label{eq:diss}
\end{equation}
Eqs. (\ref{eq:crex}) are conservative when $\alpha_X = \alpha_Y = 0$ and the
time average of a trajectory is the Nash equilibrium of the game specified by
$A$ and $B$, if a limit set exists in the interior of
$\Delta_X \times \Delta_Y$\footnote{Cf. P. Schuster et al, Biol.
Cybern. {\bf 40}, 1 (1981).}. Moreover, if the game is zero-sum, the dynamics
are Hamiltonian in ${\bf U}$ with 
\begin{eqnarray}
  H & = & -(\Sigma_j x^*_ju_j+\Sigma_j y^*_jv_j) \\ \nonumber
	& + & \log(1+\Sigma_j e^{u_j})+\log(1+\Sigma_j e^{v_j}) ~,
\end{eqnarray}
where $({\bf x}^*, {\bf y}^*)$ is an interior Nash equilibrium \cite{Hof96}.

To illustrate the dynamical-systems analysis of learning in multiagent systems
using the above framework, we now analyze the behavior of the two-person
rock-scissors-paper interaction \footnote{Such interactions are observed in
natural social and biological communities; cf. B. Kerr et al,
Nature {\bf 418}, 171 (2002).}. This
familiar game describes a nontransitive
three-sided competition: rock beats scissors, scissors beats paper, and paper
beats rock. The reward structure (environment) is given by:
\begin{equation}
A=\left[
  \begin{array}{ccc}
    \epsilon_X & 1 & -1\\
    -1 & \epsilon_X & 1\\
    1 & -1 & \epsilon_X\\
  \end{array}
  \right] ~{\rm and}~
B=\left[
  \begin{array}{ccc}
    \epsilon_Y & 1 & -1\\
    -1 & \epsilon_Y & 1\\
    1 & -1 & \epsilon_Y\\
  \end{array}
  \right] ~, 
\label{RSPGame}
\end{equation}
where $\epsilon_X, \epsilon_Y \in[-1.0,1.0]$ are the rewards for ties.
The mixed Nash equilibrium is $x^*_i=y^*_i=1/3, (i = 1,2,3)$---the centers
of $\Delta_X$ and $\Delta_Y$. If $\epsilon_X = -\epsilon_Y$, the game
is zero-sum.

In the special case of perfect memory ($\alpha_X = \alpha_Y = 0$) and with
equal learning sensitivity ($\beta_X = \beta_Y$), the linear
version (Eqs. (\ref{eq:crex})) of our model (Eq. (\ref{eq:crex1})) reduces
to multipopulation replicator equations \cite{Tay79}: 
\begin{equation}
\frac{\dot{x}_i}{x_i} =
  \left[(A{\bf y})_i-{\bf x}\cdot A{\bf y} \right]
~{\rm and}~
\frac{\dot{y}_i}{y_i} =
  \left[(B{\bf x})_i-{\bf y}\cdot B{\bf x} \right] ~. 
\label{eq:conx}
\end{equation}

\begin{figure}
  \begin{center}
    \leavevmode
	\includegraphics[scale=0.26]{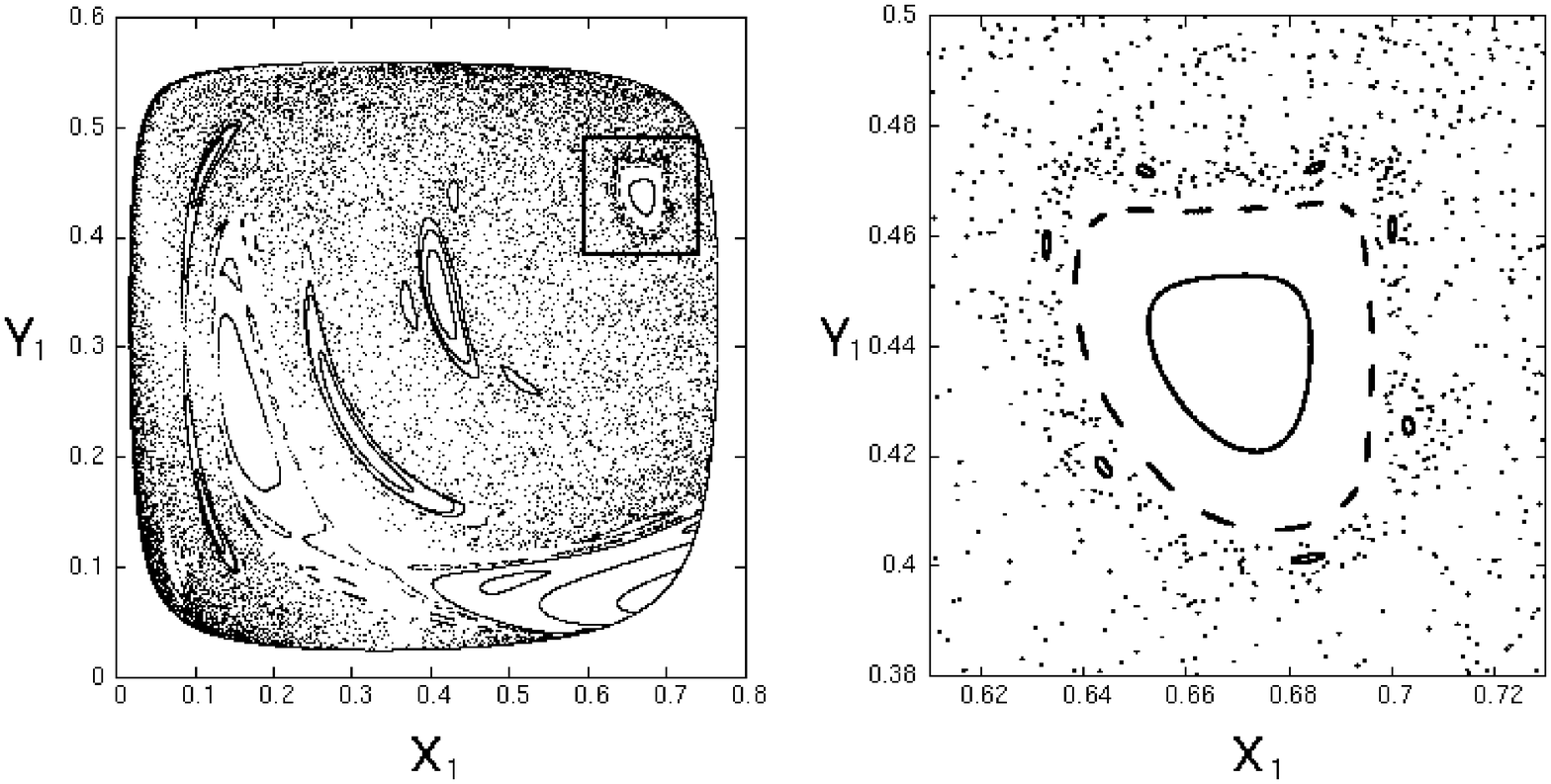}
	\includegraphics[scale=0.26]{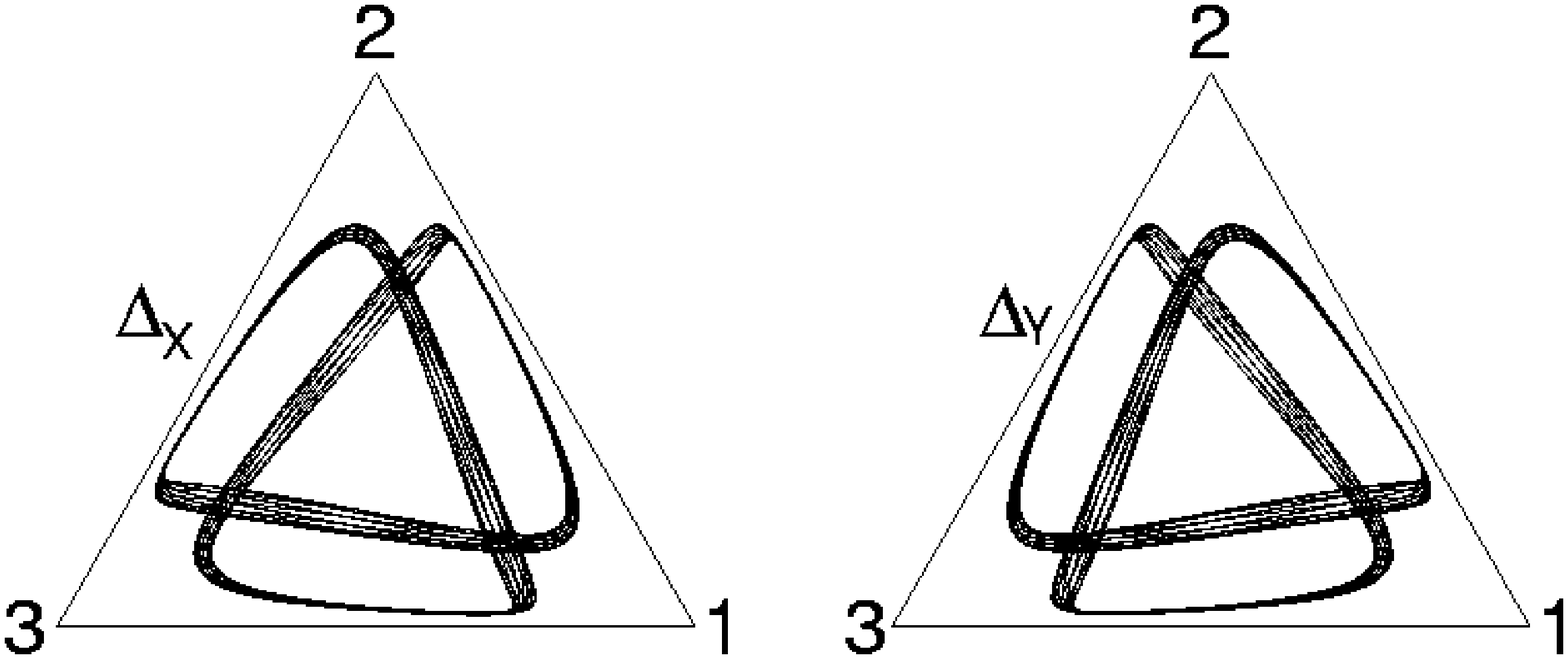}
	\includegraphics[scale=0.26]{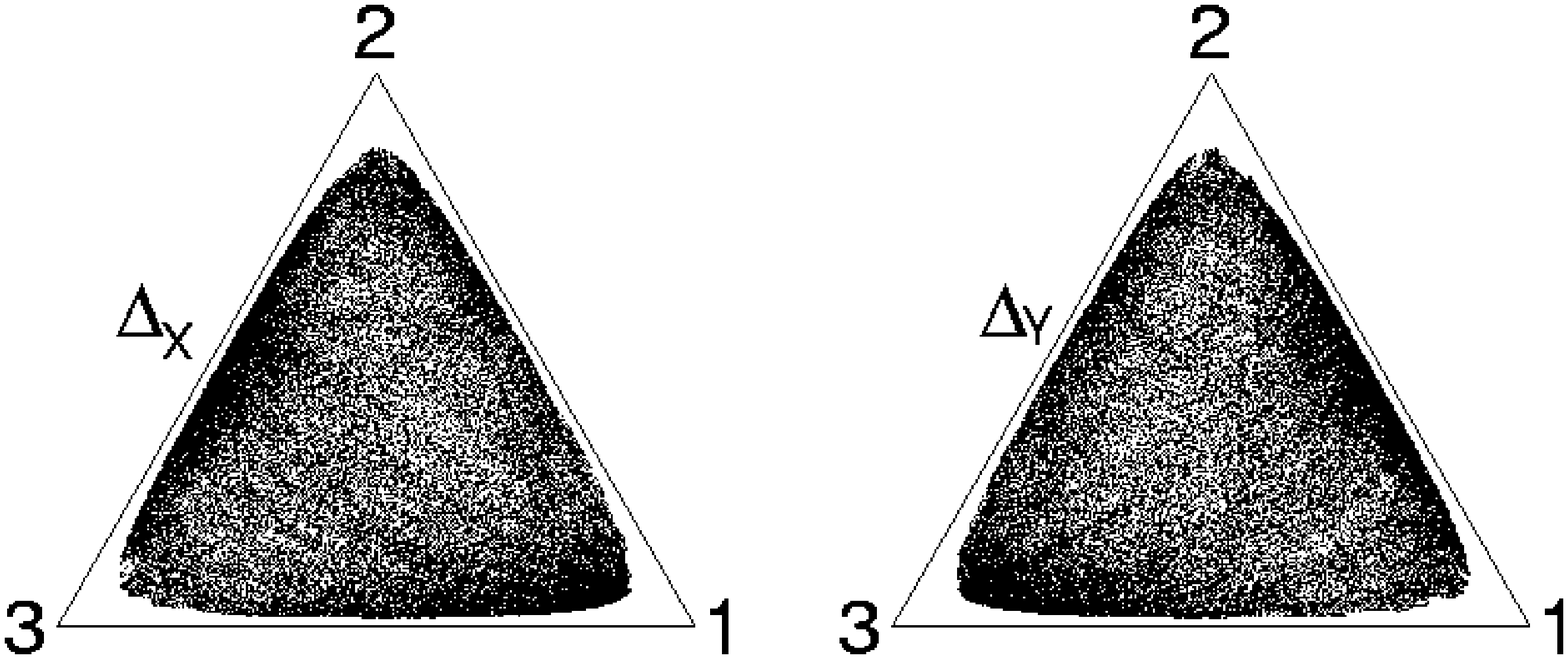}
  \end{center}
\caption{Quasiperiodic tori and chaos:
  $\epsilon_X = - \epsilon_Y = 0.5$, $\alpha_X = \alpha_Y = 0$, and
  $\beta_X = \beta_Y$. We give a 
  Poincar\'e section (top) on the hyperplane defined by $\dot{u}_1 = 0$
  and $\dot{v}_1 > 0$; that is, in the $({\bf x},{\bf y})$ space:
  $(3 + \epsilon_X) y_1 + (3 - \epsilon_X) y_2 - 2 = 0$ and
  $(3 + \epsilon_Y) x_1 + (3 - \epsilon_Y) x_2 - 2 < 0$.
  There are 23 randomly selected initial conditions with energies
  $H = -1/3(u_1+u_2+v_1+v_2)+\log(1+e^{u_1}+e^{u_2})
  +\log(1+e^{v_1}+e^{v_2})= 2.941693$, which surface 
  forms the outer border of $H\le 2.941693$. Two rows (bottom): Representative
  trajectories, simulated with a $4$th-order symplectic integrator \cite{Yos90},
  starting from initial conditions within the Poincar\'e section. The upper
  simplices show a torus in the section's upper right corner; see the enlarged
  section at the upper right. The initial condition is
  $({\bf x},{\bf y}) = (0.3, 0.054196, 0.645804,0.1,0.2,0.7)$. The lower
  simplices are an example of a chaotic trajectory passing through the regions
  in the section that are a scatter of dots; the initial condition is
  $({\bf x},{\bf y}) = (0.05, 0.35, 0.6, 0.1, 0.2, 0.7)$.
  }
\label{fig:pcs}
\end{figure}

The game-theoretic behavior in this case with rock-scissors-paper
interactions (Eqs. (\ref{RSPGame})) was investigated in \cite{Sat02}.
Here, before contrasting our more general setting, 
we briefly recall the behavior in these special cases, 
noting several additional results.

Figure \ref{fig:pcs} shows Poincar\'e sections of Eqs. (\ref{eq:conx})'s
trajectories on the hyperplane $(\dot{u}_1 = 0,\dot{v}_1>0)$ and 
representative trajectories in the individual agent simplices $\Delta_X$ and
$\Delta_Y$. When $\epsilon_X = -\epsilon_Y = 0.0$, we expect the system
to be integrable and only quasiperiodic tori should exist. 
Otherwise, $\epsilon_X = -\epsilon_Y > 0.0$, Hamiltonian chaos can
occur with positive-negative pairs of Lyapunov exponents \cite{Sat02}. 
The dynamics is very rich, there are infinitely many distinct
behaviors near the unstable fixed point at the center---the classical Nash
equilibrium---and a periodic orbit arbitrarily close to any chaotic one. 
Moreover, when the game is not zero-sum ($\epsilon_X \neq \epsilon_Y$),
transients to heteroclinic cycles are observed \cite{Sat02}: On the one
hand, there are intermittent behaviors in which the time spent near pure
strategies (the simplicial vertices) increases subexponentially with
$\epsilon_X+\epsilon_Y<0$ and, on the other hand, with
$\epsilon_X+\epsilon_Y>0$, chaotic transients persist; cf.
\cite{Cha95}. 

\begin{figure}
\begin{center}
	\leavevmode
	\includegraphics[scale=0.26]{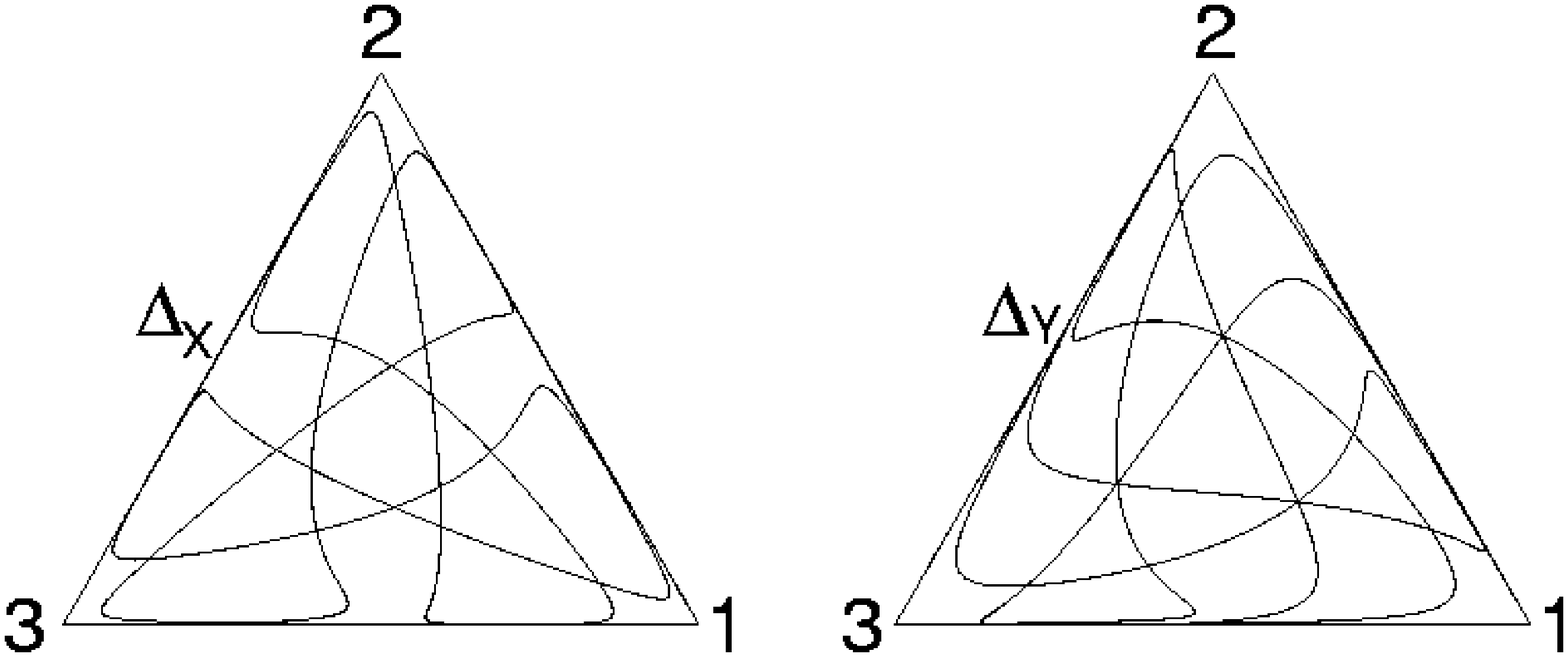}
	\includegraphics[scale=0.26]{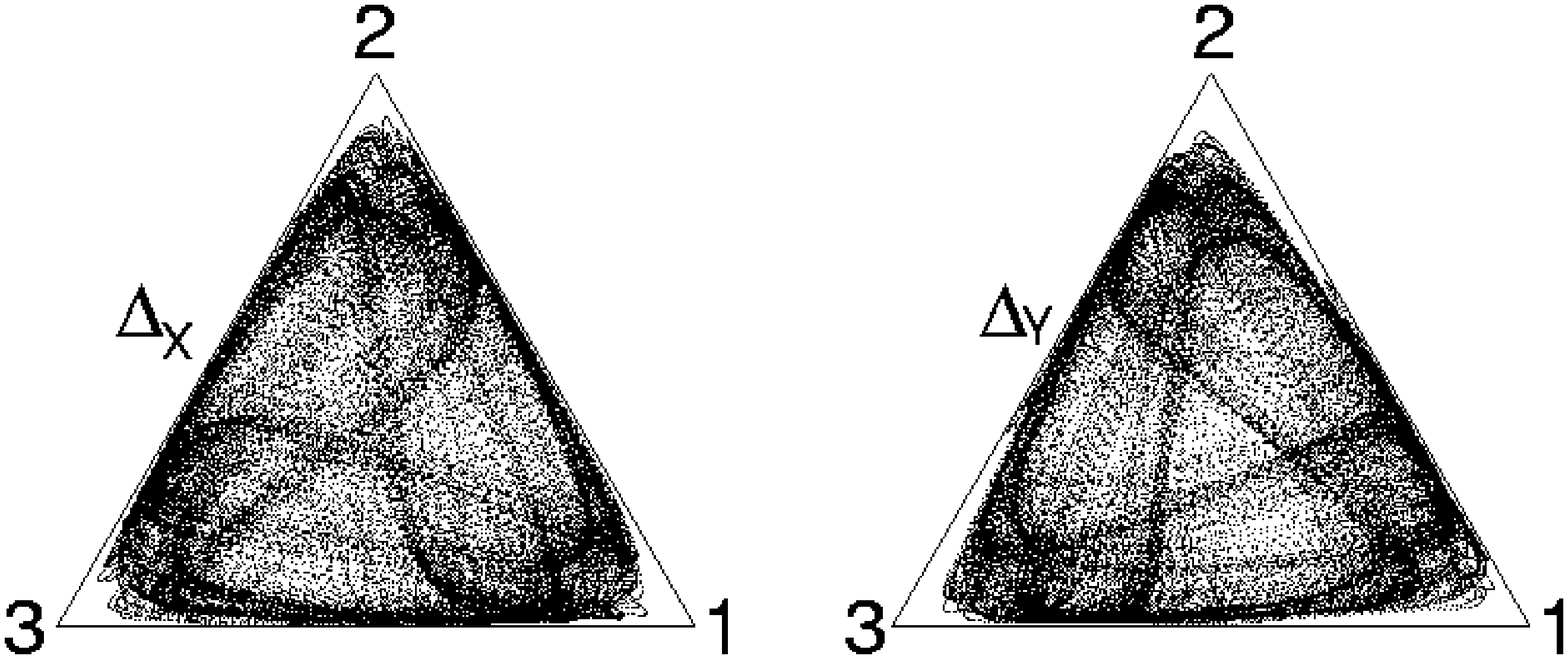}
\end{center}
\caption{Limit cycle (top: $\epsilon_Y = 0.025$) and chaotic attractors
  (bottom: $\epsilon_Y = -0.365$), with $\epsilon_X = 0.5$,
  $\alpha_X = \alpha_y = 0.01$, and $\beta_X = \beta_Y$.
  }
\label{fig:att}
\end{figure}

\begin{figure}
\begin{center}
\leavevmode
	\includegraphics[scale=0.34]{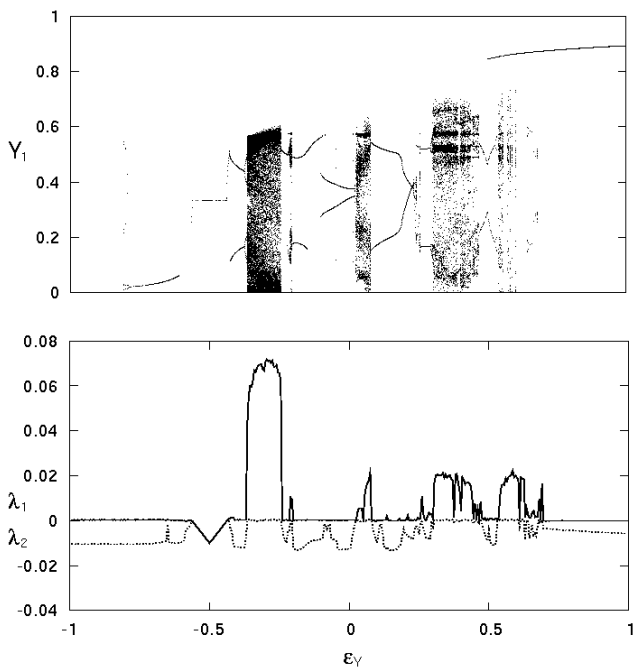}
\end{center}
\caption{Bifurcation diagram (top) of dissipative (learning with memory loss)
  dynamics projected onto coordinate $y_1$ from the Poincar\'e section
  hyperplane ($\dot{u}_1=0$, $\dot{v}_1>0$) and the largest two Lyapunov 
  exponents $\lambda_1$ and $\lambda_2$ (bottom) as a function of
  $\epsilon_Y \in [-1,1]$. Here with $\epsilon_X = 0.5$,
  $\alpha_X = \alpha_Y = 0.01$, and $\beta_X = \beta_Y$. 
  Simulations show that $\lambda_3$ and $\lambda_4$ are always negative. 
  }
\label{fig:bif}
\end{figure}

Our framework goes beyond these special cases and, generally, beyond the standard
multipopulation replicator equations (Eqs. (\ref{eq:conx})) due to its accounting
for the effects of individual and collective learning and since the 
reward structure and the learning rules need not lead to 
linear interactions.  
For example, if the memory decay rates ($\alpha_X$ and $\alpha_Y$) 
are positive, the system becomes dissipative and exhibits 
limit cycles and chaotic attractors;
see Fig. \ref{fig:att}. Figure \ref{fig:bif} (top) shows a diverse range of
bifurcations as a function of $\epsilon_Y$: dynamics on the hyperplane
($\dot{u}_1=0$, $\dot{v}_1>0$) projected onto $y_1$. When the game is nearly 
zero-sum, agents can reach the stable Nash equilibrium, but chaos can also
occur, when $\epsilon_X + \epsilon_Y > 0$. Figure \ref{fig:bif} (bottom)
shows that the largest Lyapunov exponent is positive across a significant
fraction of parameter space; indicating that chaos is common. The dual aspects
of chaos, irregularity and coherence, imply that agents may behave cooperatively
or competitively (or switch between both) in the collective dynamics. Such
global behaviors ultimately derive from self-interested, myopic learning. 

Within this framework a number of extensions suggest themselves as ways to
investigate the emergence of collective behaviors. The most obvious is the
generalization to an arbitrary number of agents with an arbitrary number of
strategies and the analysis of behaviors in thermodynamic limit; see, e.g.,
\cite{Mar00} as an alternative approach. It is relatively straightforward to
develop an extension to the linear-reward version (Eqs. (\ref{eq:crex})) of
our model. For three agents $X$, $Y$, and $Z$, one obtains:
\begin{equation}
\frac{\dot{x}_i}{x_i} = \beta_X [\Sigma_{j,k} a_{ijk} y_j z_k
  - \Sigma_{j,k,l} a_{jkl} x_j y_k z_l] - \alpha_X I_i^X ~,
\end{equation}
with tensor $(A)_{ijk} = a_{ijk}$, and similarly for $Y$ and $Z$. Not
surprisingly, this is also a conservative system when the $\alpha$'s vanish.
However, extending the general collective learning equations (Eq.
(\ref{eq:crex1})) to multiple agents is challenging and so will be reported
elsewhere. 

To be relevant to applications, one also needs to develop a statistical
dynamics generalization \cite{Nimw97a} of the deterministic equations of
motion to account for finite and fluctuating numbers of agents and also
finite histories used in learning. 
Finally, another direction, especially useful if one attempts to
quantify collective function in large multiagent systems, will be
structural and information-theoretic analyses \cite{Cru89}
of local and global learning behaviors and, importantly, their
differences. Analyzing the stored information in each agent versus
that in the collective, the causal architecture of information flow
between an individual agent and the group, and how individual and
global memories are processed to sustain collective function are
projects now made possible using this framework.

We presented a dynamical-systems model of collective learning in
multiagent systems, which starts with reinforcement learning agents and
reduces to coupled replicator equations, demonstrated that individual-agent
learning induces a global game dynamics, and investigated some of the
resulting periodic, intermittent, and chaotic behaviors with simple
(linear) rock-scissors-papers game interactions.
Our model gives a macroscopic description of a network of 
learning agents that can be straightforwardly extended 
to model a large number of heterogeneous agents 
in fluctuating environments. Since deterministic chaos occurs even in
this simple setting, one expects that in high-dimensional and heterogeneous
populations typical of multiagent systems intrinsic unpredictability will
become a dominant collective behavior. Sustaining useful collective function
in multiagent systems becomes an even more compelling question 
in light of these results.

The authors thank J. D. Farmer, E. Akiyama, P. Patelli, and C. Shalizi.
This work was supported at the Santa Fe Institute under the Network Dynamics
Program by Intel Corporation and under DARPA agreement F30602-00-2-0583.
YS's participation was supported by the Postdoctoral Researchers
Program at RIKEN.

\bibliography{credlmas}

\end{document}